\documentclass[twocolumn,english,aps,prl,superscriptaddress]{revtex4}
\usepackage[T1]{fontenc}
\usepackage[latin9]{inputenc}
\usepackage{graphicx}
\usepackage{amssymb}
\usepackage{babel}

\begin{document}

\title{Orbital coupling and superconductivity in the iron pnictides}
\author{ Junhua Zhang}
\affiliation{Department of Physics and Astronomy and Ames Laboratory, Iowa State
University, Ames, Iowa 50011, USA }
\author{ Rastko Sknepnek}
\affiliation{Department of Physics and Astronomy and Ames Laboratory, Iowa State
University, Ames, Iowa 50011, USA }
\author{Rafael M. Fernandes}
\affiliation{Department of Physics and Astronomy and Ames Laboratory, Iowa State
University, Ames, Iowa 50011, USA }
\author{ Jörg Schmalian}
\affiliation{Department of Physics and Astronomy and Ames Laboratory, Iowa State
University, Ames, Iowa 50011, USA }
\date{\today}

\begin{abstract}
We demonstrate that strong inter-orbital interaction is very efficient to
achieve superconductivity due to magnetic fluctuations in the iron
pnictides. Fermi surface states that are coupled by the antiferromagnetic
wave vector are often of different orbital nature, causing pair-hopping
interactions between distinct Fe-3d orbitals to become important. Performing
a self-consistent FLEX calculation below $T_{c}$ we determine the
superconducting order parameter as function of intra- and inter-orbital
couplings. We find an $s^{\pm}$-pairing state with $T_{c}\simeq80\mathrm{K}$
for realistic parameters.
\end{abstract}

\maketitle

High superconducting transition temperatures and the proximity to
antiferromagnetic order strongly suggest an electronic pairing mechanism in
the FeAs systems\cite{Kamihara08}. The vicinity to a spin density wave
instability with paramagnons as dominant collective mode is key for spin
fluctuation induced superconductivity, where pairing is the result of
paramagnon exchange. However, superconductivity in the iron pnictides occurs
not only in the immediate vicinity of the magnetically ordered state and the
viability of spin fluctuation induced pairing becomes an issue that requires
a quantitative analysis. In addition, multi-orbital effects of the Fe-$3d$
bands with a filling of approximately six electrons per Fe-site add to the
complexity of these systems: Electronic structure calculations\cite%
{singh08,Mazin08} yield two sets of Fermi surface sheets, one around the
center of the Brillouin zone ($\Gamma $-point) and the other around the $M$%
-point, shifted from $\Gamma $ by the magnetic ordering vector $\mathbf{Q}$%
\cite{DeLaCruz08}. Inter-band scattering of electrons has been proposed to
lead to unconventional pairing\cite%
{Mazin08,Kuroki08,Chubukov08,Cvetkovic09,Ikeda08,Seo08,Yao08,Wang09,Sknepnek09,Graser09}%
. While for certain parameters other solutions exist\cite%
{Seo08,Yao08,Graser09}, inter-band coupling tends to support the $s^{\pm }$%
-pairing state where the gap functions on the two Fermi surface sheets have
opposite sign.

\begin{figure}[tbp]
\begin{centering}
\begin{minipage}[t]{0.48\columnwidth}%
\begin{center}
\includegraphics[width=1\columnwidth]{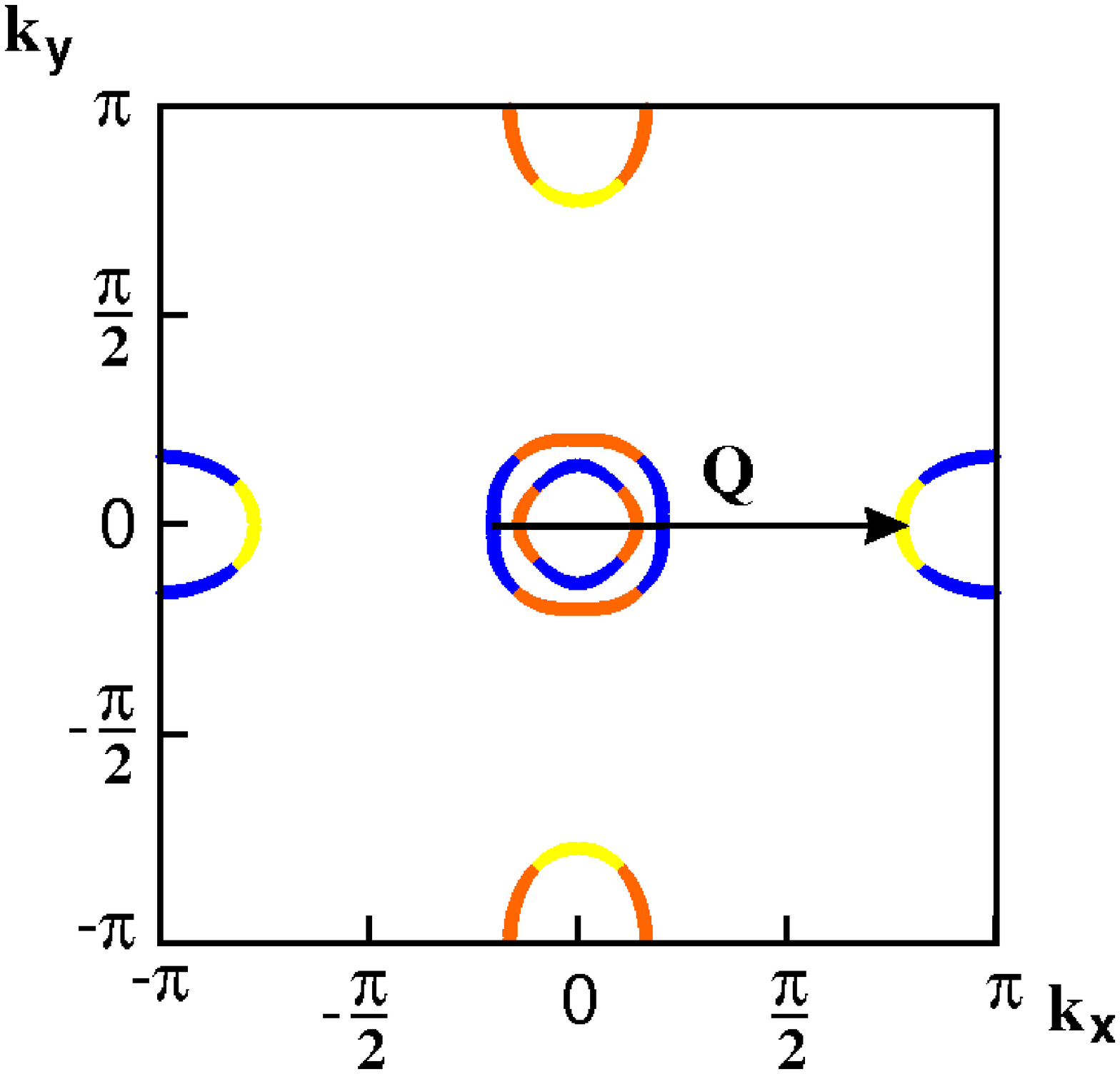}
\par\end{center}

(a)%
\end{minipage}%
\begin{minipage}[t]{0.48\columnwidth}%
\begin{center}
\includegraphics[width=1\columnwidth]{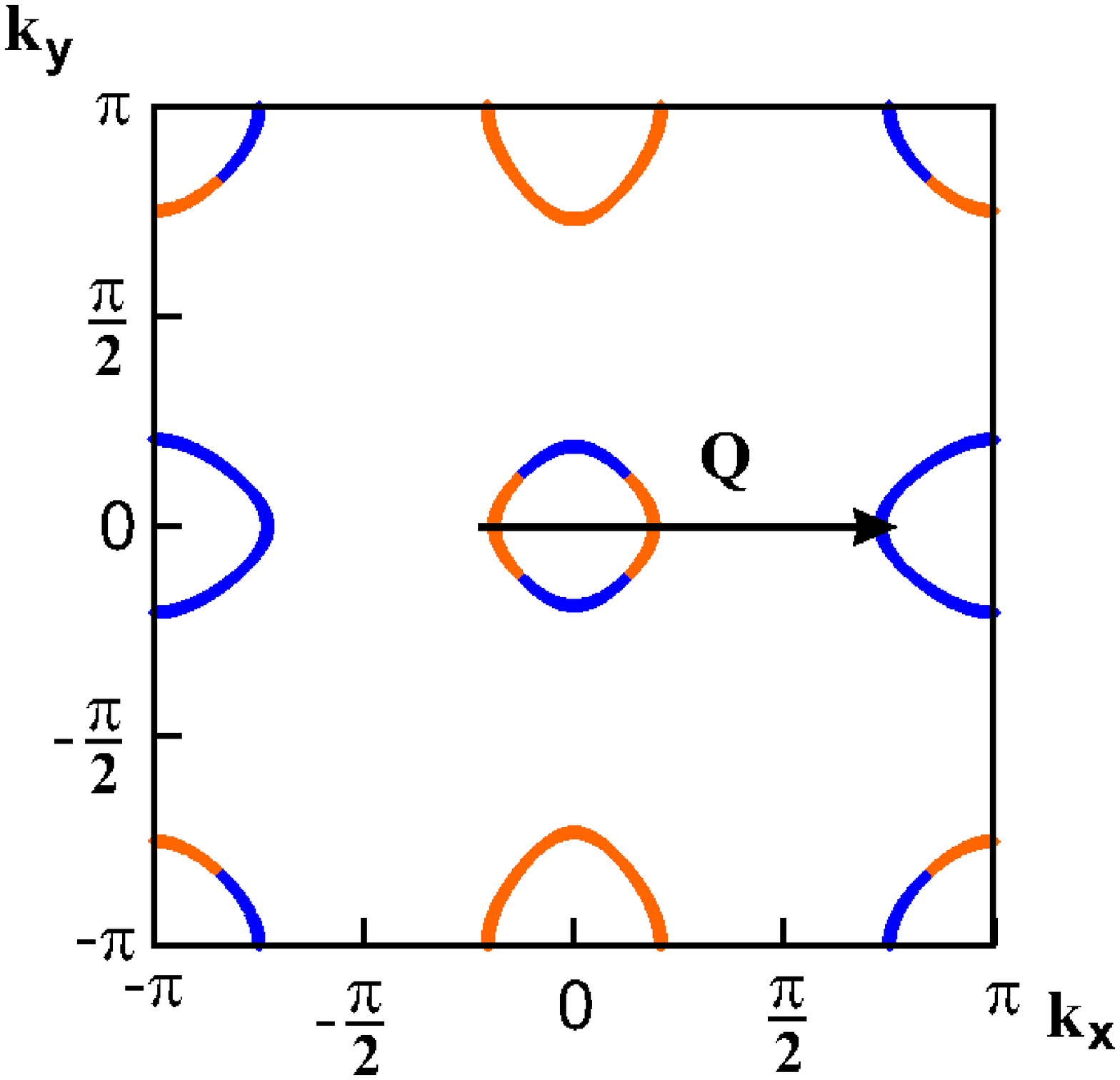}
\par\end{center}

(b)%
\end{minipage}
\par\end{centering}
\caption{(Color online) Fermi surface of the five-orbital (a) and two-orbital (b)
tight binding model of the Fe-$3d$ states in the unfolded Brillouin zone
(one iron atom per unit cell). Colors indicate the dominant orbital that
contributes to the bands: $xz$ (orange/gray), $yz$ (blue/dark gray), and $xy$ (yellow/light gray).
The antiferromagnetic vector $\mathbf{Q}=\left(\protect\pi,0\right)$ mostly
connects states dominated by \emph{different} orbitals. The tight binding
parameters are from Ref.\protect\cite{Graser09} for panel (a) and Ref.%
\protect\cite{Sknepnek09} for panel (b).}
\label{fig1}
\end{figure}

Crucial for all scenarios based upon inter-band scattering is that states $%
\left\vert \psi _{\Gamma ,\mathbf{k}}\right\rangle $ on one Fermi surface
are coupled to states $\left\vert \psi _{M,\mathbf{k+Q}}\right\rangle $ on
another Fermi surface, and vice versa. The natural starting point to
describe electron-electron interactions in transition metals is however not
in terms of bands, but rather in terms of local orbitals $\left\vert
a\right\rangle $. Here $a=xz$, $yz$, $xy$, $x^{2}-y^{2}$ and $3z^{2}-r^{2}$
refers to the Fe-$3d$ orbitals, with intra- and inter-orbital direct Coulomb
interactions, $U$ and $U^{\prime }$, as well as Hund's rule coupling $J_{H}$
and inter-orbital pair hopping $J^{\prime }$. The importance of orbital
effects in the iron pnictides was also stressed in Ref.\cite{Galitski09}. As
we will see below, the dominant effective spin-fluctuation induced pairing
interaction in a multi-orbital system is of the pair-hopping form: 
\begin{equation}
H_{\mathrm{pair}}=\sum_{\mathbf{k,k}^{\prime };a,b}W_{\mathbf{k,k}^{\prime
}}^{ab}d_{\mathbf{k}a\uparrow }^{\dagger }d_{-\mathbf{k}a\downarrow
}^{\dagger }d_{-\mathbf{k}^{\prime }b\downarrow }d_{\mathbf{k}^{\prime
}b\uparrow }.  \label{pairhop}
\end{equation}%
A pair of electrons in orbital $b$ is scattered into a pair in orbital $a$.
For $a=b$ we consider intra-orbital pairing interactions, while $a\neq b$
corresponds to an inter-orbital pairing interaction. In both cases, Cooper pairs are predominantly made up of electrons in the same orbital: $%
\left\langle d_{-\mathbf{k}a\downarrow }d_{\mathbf{k}a\uparrow
}\right\rangle \neq 0$. In the band picture this yields the inter-band
pairing interaction 
\begin{equation}
W_{\mathbf{k,k}^{\prime }}^{\Gamma ,M}\simeq \sum_{ab}\left\langle \psi
_{\Gamma ,\mathbf{k}}|a\right\rangle ^{2}W_{\mathbf{k,k}^{\prime
}}^{ab}\left\langle b|\psi _{M,\mathbf{k}^{\prime }}\right\rangle ^{2}.
\label{band}
\end{equation}%
The dominant momentum transfer in the spin fluctuation approach is of course 
$\mathbf{k-k}^{\prime }=\mathbf{Q}$. It is interesting to observe that
electronic structure calculations show that $\left\vert \psi _{\Gamma ,%
\mathbf{k}}\right\rangle $ and $\left\vert \psi _{M,\mathbf{k+Q}%
}\right\rangle $ are often dominated by \emph{different} orbitals. For
example, if $\left\langle xz|\psi _{\Gamma ,\mathbf{k}}\right\rangle $ is
large, it holds that $\left\langle xz|\psi _{M,\mathbf{k+Q}}\right\rangle $
for the same $\mathbf{k}$ is small, while $\left\langle xy|\psi _{M,\mathbf{%
k+Q}}\right\rangle $ might be large. In Fig. \ref{fig1}a we illustrate this
effect where distinct colors refer to the dominant orbitals on the Fermi
surface. We used the tight binding parametrization of the five band model of
Ref.\cite{Graser09}, where a similar plot was presented. The three dominant
orbitals on the Fermi surface are $xz$, $yz$, and $xy$. Connecting a Fermi
surface point by $\mathbf{Q=}\left( \pi ,0\right) $ or $\left( 0,\pi \right) 
$ leads in most cases to a different orbital. Thus, the orbital composition
of the wave function at the Fermi surface \emph{frustrates} intra-orbital
pairing. In other words, inter-band but intra-orbital scattering of spin
fluctuations ($\propto W_{\mathbf{k,k}^{\prime }}^{aa}$) provides a less
efficient pairing glue, if compared to inter-orbital scattering $\propto W_{%
\mathbf{k,k}^{\prime }}^{ab}$ ($a\neq b$) of equal size. It is crucial to
determine under what conditions collective paramagnons with strong
inter-orbital pair-hopping exist.

In this Rapid Communication we solve the two-orbital many-body problem in
the superconducting state for varying intra-orbital ($U$) and inter-orbital (%
$U^{\prime }$, $J\equiv J_{H}=$ $J^{\prime }$) couplings within the
self-consistent fluctuation exchange (FLEX) approximation\cite{Bickers89}.
We obtain superconductivity with $s^{\pm }$-pairing. The superconducting
order parameter is determined self-consistently and vanishes at $T_{c}\simeq
80\mathrm{K}$. We demonstrate that strong collective inter-orbital
spin-fluctuations are efficient to increase superconductivity. To solve the
FLEX\ equation on the imaginary frequency axis for a lattice of $N=32\times
32$ sites and at temperatures as low as $T\simeq 10\mathrm{K}$ we require $%
2^{13}=8192$ Matsubara frequencies. At the moment, this restricts our
analysis to consider only two orbitals. In Fig. \ref{fig1}b we show the
Fermi surface of a two band model with $d_{xz}$ and $d_{yz}$ orbitals. The
mentioned frustration of intra-orbital pairing is less pronounced for this
simplified model. Yet, the phase space for inter-orbital pairing
interactions is still larger compared to intra-orbital interactions.

\emph{The model:} We consider the Hamiltonian 
\begin{eqnarray}
H &=&\sum_{\mathbf{k},ab,\sigma }\varepsilon _{\mathbf{k}}^{ab}d_{\mathbf{k}%
a\sigma }^{\dagger }d_{\mathbf{k}b\sigma }-J_{H}\sum_{i,a>b}\left( 2\mathbf{s%
}_{ia}\cdot \mathbf{s}_{ib}+\frac{1}{2}n_{ia}n_{ib}\right)   \nonumber \\
&+&U\sum_{i,a}n_{ia\uparrow }n_{ia\downarrow }+U^{\prime
}\sum_{i,a>b}n_{ia}n_{ib}+J^{\prime }\sum_{i,a\neq b}b_{ia}^{\dagger }b_{ib},
\label{eq:1}
\end{eqnarray}%
where $n_{ia\sigma }=d_{ia\sigma }^{\dagger }d_{ia\sigma }$ is the
occupation of the orbital $a$ with spin $\sigma $ at site $i$ and $%
n_{ia}=\sum_{\sigma }n_{ia\sigma }$.  $\mathbf{s}_{ia}=\frac{1}{2}%
\sum_{\sigma \sigma ^{\prime }}d_{ia\sigma }^{\dagger }\mathbf{\sigma }%
_{\sigma \sigma ^{\prime }}d_{ia\sigma ^{\prime }}$ is the electron spin and 
$b_{ia}^{\dagger }=d_{ia\uparrow }^{\dagger }d_{ia\downarrow }^{\dagger }$
the pair creation operator, respectively. For the tight binding band
structure we use $\varepsilon _{\mathbf{k}}^{xy}=\varepsilon _{\mathbf{k}%
}^{yx}=-4t_{4}\sin k_{x}\sin k_{y}$ and $\varepsilon _{\mathbf{k}%
}^{aa}=-2t_{1}\cos k_{a}-2t_{2}\cos k_{\overline{a}}-4t_{3}\cos k_{x}\cos
k_{y}-\mu $, where $a=x$ ($y$) stands for $xz$ ($yz$) orbital as well as the
momentum coordinate with $\overline{x}=y$ and $\overline{y}=x$. We use $%
t_{1}=-0.33\ \mathrm{eV}$, $t_{2}=0.385\ \mathrm{eV}$, $t_{3}=-0.234\ 
\mathrm{eV}$, and $t_{4}=-0.26\,\mathrm{eV}$ of Ref. \cite{Sknepnek09}. Our
results were obtained for a filling of $n=1.88$ electrons per site,
corresponding to moderate hole doping. The filling of a subset of bands is
primarily determined to reproduce realistic Fermi surface geometries and
yields commensurate magnetic fluctuations. 

\emph{The multi-orbital fluctuation exchange approach:} The FLEX equations
for a multi-orbital problem are given in Ref.\cite{Takimoto04}. In the
normal state one obtains the single particle self energy $\Sigma _{k}^{ab}$
which yields the single particle propagator $G_{k}^{ab}$. Here $k=\left( 
\mathbf{k,}i\omega _{n}\right) $ stands for the momentum vector $\mathbf{k}$
and the Matsubara frequency $\omega _{n}=\left( 2n+1\right) \pi T$ with
temperature $T$. As it is important for our subsequent discussion we
summarize the key equations that occur in the superconducting state and
determine the anomalous self energy: 
\begin{equation}
\Phi _{k}^{ab}=\sum_{k^{\prime }}\sum_{cd}\Gamma _{k-k^{\prime
}}^{ac,db}F_{k^{\prime }}^{cd}.  \label{sc1}
\end{equation}%
This equation is the strong coupling version of the gap-equation. $%
\sum_{k}\ldots =\frac{T}{N}\sum_{\mathbf{k},n}\ldots $ stands for the
summation over momenta and Matsubara frequencies. $F_{k}^{ab}$ is the
anomalous Green's function, that determines the Cooper pair expectation
value $\left\langle d_{\mathbf{k}a\mathbf{\uparrow }}d_{-\mathbf{k}b\mathbf{%
\downarrow }}\right\rangle =T\sum_{n}F_{k}^{ab}$. Furthermore, $\Gamma
_{q}^{ac,db}$ is the dynamic pairing interaction that depends on momentum,
frequency and the orbital states involved,  where $q=\left( \mathbf{q,}i\nu
_{n}\right) $ with $\nu _{n}=2n\pi T$. Introducing the two-particle quantum
numbers $A=\left( a,c\right) $ and $B=\left( d,b\right) $ that label the
rows and columns of two-particle states, the interaction, $\Gamma
_{q}^{A,B}=\Gamma _{q}^{ac,db}$ becomes a $m^{2}\times m^{2}$-dimensional
symmetric operator $\widehat{\Gamma }_{q}$, where $m$ is the number of
orbitals. \ It is now straightforward to sum particle hole ladder and bubble
diagrams. It follows: 
\begin{equation}
\widehat{\Gamma }_{q}=\frac{3}{2}\widehat{V}_{s,q}+\frac{1}{2}\widehat{V}%
_{c,q}+\widehat{V}_{HF},  \label{ns2}
\end{equation}%
where interactions in the spin and charge channel are: 
\begin{eqnarray}
\widehat{V}_{s(c),q} &=&\pm \widehat{U}_{s\left( c\right) }\left( 1\mp 
\widehat{\chi }_{s\left( c\right) ,q}\widehat{U}_{s\left( c\right) }\right)
^{-1}\widehat{\chi }_{s\left( c\right) ,q}\widehat{U}_{s\left( c\right) } 
\nonumber \\
&&-\frac{1}{4}\widehat{U}_{s\left( c\right) }\left( \widehat{\chi }_{s,q}-%
\widehat{\chi }_{c,q}\right) \widehat{U}_{s\left( c\right) }.  \label{sc3}
\end{eqnarray}%
$\widehat{U}_{s}$ and $\widehat{U}_{c}$ are also $m^{2}\times m^{2}$%
-dimensional matrices of the interaction in the spin and charge channel,
respectively. Close to a magnetic instability, the dominant contribution to $%
\widehat{\Gamma }_{q}$ comes from the spin channel $\widehat{V}_{s,q}$ due
to the Stoner enhancement $\left( 1-\widehat{\chi }_{s,q}\widehat{U}%
_{s}\right) ^{-1}$. For the interaction matrix in the spin sector holds $%
U_{s}^{aa,aa}=U$, while for $a\neq b$ holds $U_{s}^{ab,ab}=U^{\prime }$, $%
U_{s}^{ab,ba}=J^{\prime }$ and $U_{s}^{aa,bb}=J_{H}$. The Hartree-Fock
contribution $\widehat{V}_{HF}=(\widehat{U}_{s}+\widehat{U}_{c})/2$ is
suppressing superconductivity, an effect caused by the repulsive direct
Coulomb interaction. We find that the impact of direct Coulomb interaction
is strongly reduced in the $s^{\pm }$ state with small, but finite, average $%
\left\langle d_{ia\mathbf{\uparrow }}d_{ia\mathbf{\downarrow }}\right\rangle 
$ for local Cooper pairing. For a discussion of this Coulomb avoidance see
Ref. \cite{Mazin09}. Finally, the irreducible particle hole bubble $\widehat{%
\chi }_{s\left( c\right) ,q}$ is determined by normal and anomalous Green's
functions: $\chi _{s\left( c\right) ,q}^{ab,cd}=-\sum_{k}\left(
G_{k+q}^{ac}G_{k}^{db}\pm F_{k+q}^{ad}F_{k}^{cb}{}^{\ast }\right) $,
assuming time reversal invariance and singlet pairing. We solved the set of
coupled FLEX equations self consistently in the superconducting state.

\begin{figure}[tbp]
\begin{centering}
 \includegraphics[width=0.675\columnwidth,angle=270]{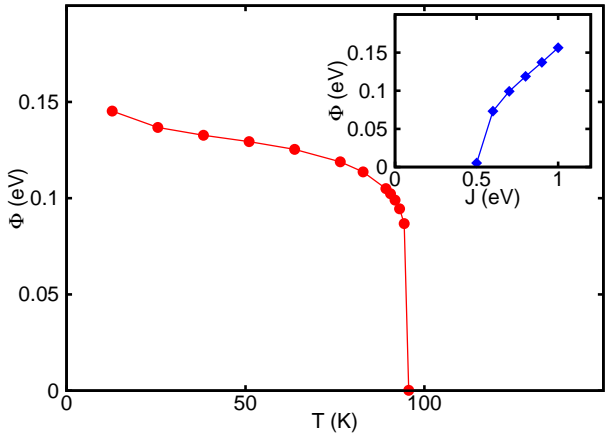}
\par\end{centering}
\caption{(Color online) Temperature dependence of the anomalous self energy $%
\Phi_{\mathbf{k=0},\protect\pi T}^{xx}$, proportional to the superconducting
order parameter, for $U=1.5\mathrm{eV}$, $U^{\prime}=1.2\mathrm{eV}$, and $%
J=0.8\mathrm{eV}$ . The inset shows the increase of $\Phi_{k}^{xx}$ with
increasing inter-orbital coupling $J$ at $T=60\mathrm{K}$.}
\label{fig2}
\end{figure}

\begin{figure}[tbp]
\begin{centering}
 \includegraphics[width=0.95\columnwidth]{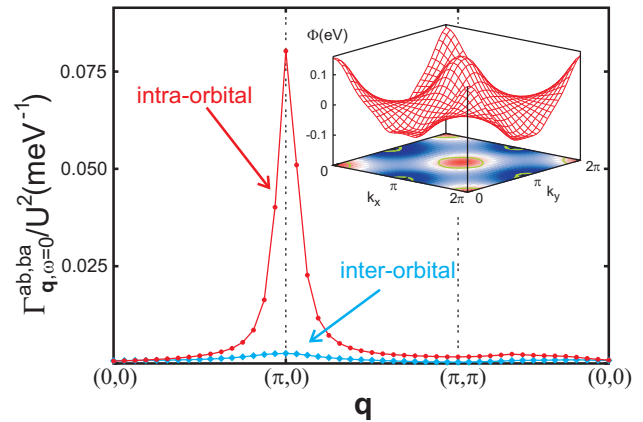}
\par\end{centering}
\caption{(Color online) Intra-orbital and inter-orbital pairing interaction
at zero frequency as function of momentum for $U=1.5$eV, $U^{\prime}=1$eV
and $J=1$eV at a temperature of $T=70$K along with the self-consistently
determined anomalous self energy $\Phi_{\mathbf{k,}\protect\pi T}^{xx}$ (%
\emph{inset}) that determines the sign and momentum dependence of the
superconducting gap. The pairing interaction is peaked for momenta $\mathbf{%
Q=}\left(\protect\pi,0\right)$ and $\left(0,\protect\pi\right)$. The
anomalous self energy corresponds to $s^{\pm}$-pairing with opposite sign of
the gap on Fermi surface sheets around $\Gamma$ and $M$. Inter-orbital
pairing is significantly smaller. }
\label{fig3}
\end{figure}

\begin{figure}[tbp]
\begin{centering}
 \includegraphics[width=0.95\columnwidth]{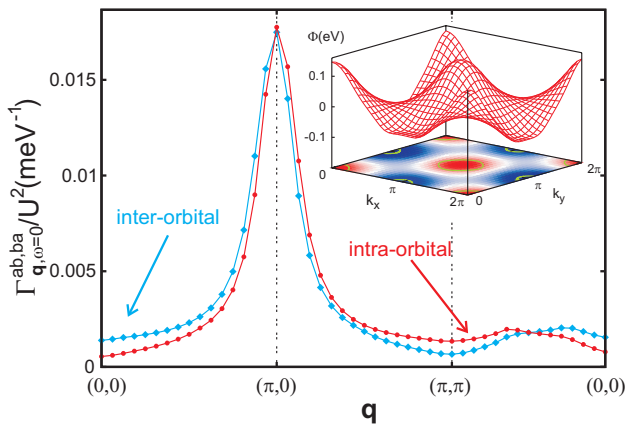}
\par\end{centering}
\caption{(Color online) Same as Fig. \protect\ref{fig3} but for parameters $%
U=U^{\prime}=1.5\mathrm{eV}$ and $J=1\mathrm{eV}$ with same temperature $T=70%
\mathrm{K}$. Now intra- and inter-orbital couplings are of comparable
size, leading to the same superconducting order parameter as for Fig. 
\protect\ref{fig3}, however for much smaller pairing strength, associated
with moderate antiferromagnetic fluctuations (see the different scale
compared to Fig. \protect\ref{fig3}). }
\label{fig4}
\end{figure}

\emph{The pairing state and its} $T$\emph{-dependence:} The momentum
dependence of the anomalous self energy $\Phi _{k}^{aa}$ is shown in the
insets of Figs. \ref{fig3} and \ref{fig4}. The symmetry of $\Phi _{k}^{ab}$
and of the Hamiltonian are the same, corresponding to $s$-wave pairing\cite%
{Sknepnek09}. Nevertheless, the sign of $\Phi _{k}^{aa}$ is opposite on
Fermi surface sheets around $\Gamma $ and $M$, i.e., we obtain the $s^{\pm }$
pairing state that was proposed in Ref. \cite{Mazin08}. While in general
such a state can have nodes of the gap on the Fermi surface, our solution
corresponds to a fully gapped state. For a recent discussion of the $s^{\pm }
$-state see Ref. \cite{Mazin09}. In Fig. \ref{fig2} we show the temperature
dependence of the anomalous self energy $\Phi _{k=0}^{xx}$, which is
proportional to the superconducting order parameter. The feedback of the
opening of a pairing gap onto the dynamic pairing interaction leads to the
rather rapid growth of the order parameter below $T_{c}$\cite{Monthoux94}. $%
T_{c}\simeq 80\mathrm{K}$ is indeed of the correct order of magnitude. In
the normal state the dynamics of paramagnons is overdamped $\Gamma _{\mathbf{%
Q},\omega }^{aa,aa}\sim \left( 1+|\omega |/\omega _{s}\right) ^{-1}$. For
the parameters of Fig. \ref{fig2} we find $\omega _{s}\left( T_{c}\right) =37%
\mathrm{meV}$. This energy scale is reduced compared to the typical
electronic energies because of the Stoner enhancement. It sets the scale for
the Lorentzian lineshape of  inelastic neutron scattering at $\mathbf{Q}$
above $T_{c}$. A strong pairing interaction is caused by a significant
Stoner enhancement, which is controlled by the value of the magnetic
inter-orbital coupling, $J$. This is demonstrated in the inset of Fig. \ref%
{fig2} where we show the sensitivity of superconducting order with respect
to $J$ at $T=60\mathrm{K}$.

\emph{Intra- versus inter-orbital pairing:} For the pnictides, the  pairing
vertex \ $\widehat{\Gamma }_{q}$  in Eq. (\ref{sc1}) is dominated by a few
matrix elements. We find that $\chi _{s,q}^{A,B}$ at $\mathbf{q\simeq Q}$
has comparable diagonal elements $\chi _{d}$ and somewhat smaller counter
diagonal elements $\chi _{\overline{d}}$ in two-particle space, while all
other matrix elements are negligible. It then follows from Eq. (\ref{sc3})
that the dominant matrix elements of $\widehat{\Gamma }_{q}$ are $\Gamma
_{q}^{ab,ba}$. If one interprets $\Gamma _{q}^{ab,ba}$ as effective low
energy interaction, the combination of orbital indices yields precisely the
pair-hopping form Eq. (\ref{pairhop}) with $W_{k,k^{\prime }}^{ab}=\Gamma
_{k-k^{\prime }}^{ab,ba}$. The effective Stoner enhancements for same
orbitals $a=b$ are $\Gamma _{q}^{aa,aa}\simeq \left( U+J_{H}\right) \left(
1-\left( U+J_{H}\right) \left( \chi _{d}+\chi _{\overline{d}}\right) \right)
^{-1}$ while for $a\neq b$ follows $\Gamma _{q}^{ab,ba}\simeq \left(
U^{\prime }+J^{\prime }\right) \left( 1-\left( U^{\prime }+J^{\prime
}\right) \left( \chi _{d}+\chi _{\overline{d}}\right) \right) ^{-1}$ .
Whenever $U$ is significantly larger than the other couplings, Eq. (\ref{sc1}%
) is dominated by interactions within the same orbital. As mentioned, the
intra-orbital pairing interaction is however rather inefficient. The
situation changes when we consider comparable values for the intra- and
inter-orbital Coulomb enhancements: $U+J_{H}\gtrsim U^{\prime }+J^{\prime }$,
i.e. a regime with strong orbital fluctuations. The pairing interaction is
enhanced, as the nature of the wave functions on the Fermi surface can
efficiently take advantage of coupling between distinct orbitals, see Eq. (%
\ref{band}) and Fig.1. 

The condition  $U+J_{H}\gtrsim U^{\prime }+J^{\prime }$ is at variance with
the relations $U=U^{\prime }+2J$ and $J^{\prime }=J_{H}$ that result from
the rotational symmetry of the bare Coulomb interaction\cite{Castellani78},
if combined with evidence for sizable Hund coupling\cite{Kroll08,Haule09}.
We stress however that the interaction parameters that enter an approximate
theory such as FLEX are not identical to the bare Coulomb matrix elements%
\cite{Bulut93}. FLEX ignores crucial vertex corrections and $U$, $U^{\prime }
$ and $J$ should rather be considered low energy interaction parameters that
have been renormalized by high energy excitations. Performing a
renormalization of the Coulomb interactions within a multiband version of
the Kanamori scattering matrix approach\cite{Kanamori63}, we indeed find
that $U+J_{H}\simeq U^{\prime }+J^{\prime }$\  for realistic values of the
bare Coulomb matrix elements of Fe\cite{Miyake08}. In this approach, particle-particle excitations couple states with $\mathbf{k}$ and $\mathbf{-k}$, i.e., states of same orbital nature. This reduces $U$ more strongly than $U'$, $J_H$, and $J'$. Thus, constraints due to rotational invariance do not apply
for low energy vertices that enter FLEX, and intra-orbital and inter-orbital
pairing interactions can easily be comparable. The underlying low density expansion makes the Kanamori scattering matrix renormalization a very sensible approach for the pnictides, given their near semimetallic  electronic structure.

In Figs. \ref{fig3} and \ref{fig4} we compare two different parameter sets
that yield almost the same value for $\Phi _{k}^{aa}$ at the same
temperature. The first case has a predominant intra-orbital interaction
which needs to be very large in order to achieve pairing. In the other case
the same pairing amplitude is obtained from intra- and inter-orbital pairing
interactions. $\Gamma _{q}^{ab,ba}$ in Fig. \ref{fig4} are almost one fifth
of the pure intra-band interaction in Fig. \ref{fig3}, demonstrating the
efficient role played by inter-orbital magnetic pairing interactions in the
iron pnictides. 

In summary we presented a self-consistent FLEX analysis of a two orbital model
of the FeAs systems in the superconducting state. We determined the
temperature dependence of the superconducting order parameter and showed
that $T_{c}\simeq 60-80\mathrm{K,}$ of the order of the experimental values,
are clearly possible. The pairing state is $s^{\pm }$ with opposite sign of
the gap on Fermi surface sheets around $\Gamma $ and $M$\cite{Mazin08}. In
the iron pnictides, states that are coupled by the antiferromagnetic wave
vector are often dominated by different local Fe-$3d$ orbitals. This makes a
purely intra-orbital pairing interaction quite inefficient. Inter-orbital
pairing due to antiferromagnetic fluctuations yields the same pairing amplitude for much smaller Stoner
enhancement, i.e., for more moderate values of the magnetic correlation length. We expect this effect to be even stronger in a
more realistic five-orbital description of the iron pnictides. Collective low
energy pairing interaction between like and unlike orbitals, i.e., strong
orbital fluctuations, significantly enhances \ the viability of the spin
fluctuation approach for superconductivity in the pnictides.

We are grateful to A. V. Chubukov and I. I. Mazin for helpful discussions.
This research was supported by the Ames Laboratory, operated for the U.S.
Department of Energy by Iowa State University under Contract No.
DE-AC02-07CH11358.

\end{document}